\documentclass[aps,prl,reprint,amsmath,amssymb,superscriptaddress,noeprint]{revtex4-2}
\usepackage{changes}

\usepackage{xspace}
\usepackage{titlesec}

\usepackage{amssymb,bm}
\usepackage{graphicx}
\usepackage{amsmath}
\usepackage{braket}
\usepackage{bbm}
\usepackage{units}

\newcommand{\kB}{k_\mathrm{B}}
\newcommand{\EF}{E_\mathrm{F}}
\newcommand{\kF}{k_\mathrm{F}}
\newcommand{\F}{\mathrm{F}}
\newcommand{\p}{\text{p}}
\newcommand{\B}{\text{B}}

\newcommand{\kFa}{k_\mathrm{F}a_{\uparrow\B}}
\newcommand{\Minf}{\mathcal{M}_\infty}
\newcommand{\diff}{\mathrm{d}}
\usepackage[colorlinks=true,linkcolor=blue,citecolor=blue,urlcolor=blue]{hyperref}
\usepackage{xcolor,soul}
\usepackage{lineno}

\makeatletter
\def\maketitle{
\@author@finish
\title@column\titleblock@produce
\suppressfloats[t]}

\begin{document}
\title{The Kubo-Thermalization Correspondence}

\author{Songtao Huang}
\thanks{These authors contributed equally to this work}
\affiliation{Department of Physics, Yale University, New Haven, Connecticut 06520, USA}

\author{Xingyu Li}
\thanks{These authors contributed equally to this work}
\affiliation{Institute for Advanced Study, Tsinghua University, Beijing 100084, China}

\author{Jianyi Chen}
\affiliation{Department of Physics, Yale University, New Haven, Connecticut 06520, USA}

\author{Alan Tsidilkovski}
\affiliation{Department of Physics, Yale University, New Haven, Connecticut 06520, USA}

\author{Gabriel G. T. Assumpção}
\affiliation{Department of Physics, Yale University, New Haven, Connecticut 06520, USA}

\author{Pengfei Zhang}
\affiliation{Department of Physics, Fudan University, Shanghai 200438, China}

\author{Hui Zhai}
\email{hzhai@tsinghua.edu.cn}
\affiliation{Institute for Advanced Study, Tsinghua University, Beijing 100084, China}

\author{Nir Navon}
\email{nir.navon@yale.edu}
\affiliation{Department of Physics, Yale University, New Haven, Connecticut 06520, USA}
\affiliation{Yale Quantum Institute, Yale University, New Haven, Connecticut 06520, USA}

\begin{abstract}
Quantum thermalization describes how interacting quantum systems relax toward thermal equilibrium~\cite{linden2009quantum,nandkishore2015many,eisert2015quantum}, a central problem in modern physics. Yet most experimental information on many-body systems comes from short-time transition spectroscopy, typically interpreted within Kubo’s linear-response framework~\cite{fermi1950nuclear,kubo1957statistical}. These perspectives—long-time equilibration versus short-time response—seem fundamentally disconnected. Here we establish an exact link between them: the \emph{Kubo-Thermalization correspondence}, which connects long-time thermalized magnetization under weak driving to short-time linear-response spectra for a spin coupled to a thermal bath. The correspondence holds even when the steady state differs substantially from the initial state and when each regime is individually difficult to describe theoretically~\cite{PhysRevLett.133.083403,PhysRevA.109.023302}. We experimentally confirm the correspondence using effective spin-$1/2$ impurities realized with ultracold fermions in two internal states coupled to a Fermi sea~\cite{vivanco2025strongly}. Our results provide a rare exact statement about quantum thermalization and offer a novel route to infer thermalization dynamics from equilibrium response measurements in strongly interacting quantum systems, independent of microscopic details of the system--bath coupling.
\end{abstract}

\date{\today}

\maketitle

\begin{figure}[h]
\includegraphics[width=1\columnwidth]{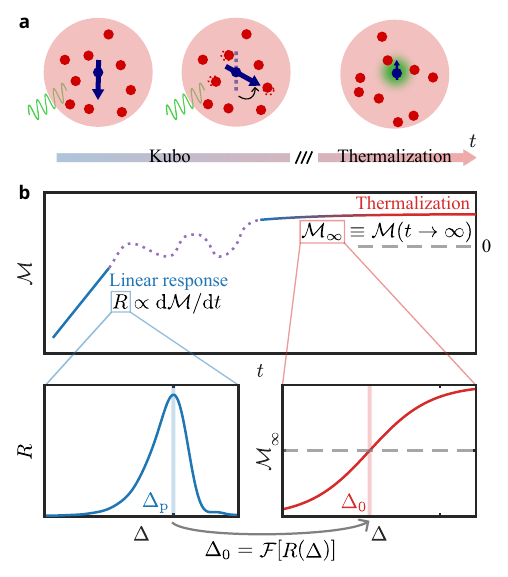}
\caption{
\textbf{Kubo-Thermalization correspondence for a driven spin coupled to a thermal bath.} (\textbf{a}) A spin-1/2 particle is immersed in a thermal bath at temperature $T$, and an external spin-flip term drives the quantum dynamics for a duration $t$ with a weak Rabi frequency $\Omega_0$. (\textbf{b}) (Top) Dynamical evolution of the magnetization $\mathcal{M}$ after initializing the spin in $\ket{\downarrow}$ and turning on a weak spin-flip field. (Bottom left) The short-time transition spectroscopy $R(\Delta)$, defined as the transition rate versus detuning, and $\Delta_\p$ is its peak position. (Bottom right) The long-time steady-state magnetization $\mathcal{M}_\infty(\Delta)$, characterized by its zero crossing $\Delta_0$ (defined as $\mathcal{M}_\infty(\Delta_0)=0$). Our central result, Eq. \eqref{correspondence}, is a rigorous functional relation $\Delta_0=\mathcal{F}[R(\Delta)]$ that connects $\Delta_0$ to the spectrum $R(\Delta)$.}
\label{Fig1}
\end{figure}

Determining the quantum dynamics of strongly correlated many-body systems is notoriously difficult: there is no universal route from the full, exponentially large Hilbert-space description to a compact, predictive theory. When such systems are driven weakly, however, Kubo’s linear-response framework and Fermi’s Golden Rule (FGR)~\cite{fermi1950nuclear,kubo1957statistical} provide a powerful simplification at short times: dynamical observables reduce to correlation functions evaluated on the initial equilibrium state. 
This viewpoint underpins much of modern spectroscopy, including ARPES~\cite{damascelli2003angle,boschini2024time}, neutron scattering~\cite{furrer2009neutron}, and Raman spectroscopy~\cite{devereaux2007inelastic}.

For longer times, the situation changes dramatically: switching on even a weak drive renders the initial equilibrium reference progressively irrelevant. Indeed, a system prepared in equilibrium for the \emph{undriven} Hamiltonian is evolved under the \emph{driven} dynamics—an effective “drive quench” that takes it out of equilibrium with respect to the new conditions; it may then thermalize to a steady state far from the initial one despite the drive being weak. It is therefore far from obvious that short-time linear-response data around the initial equilibrium should constrain, much less encode, the ensuing long-time thermalization under sustained driving.

Here we establish a direct link between the properties of a thermalized driven state and the short-time weak-drive response of a spin coupled to a bath. Importantly, this relation formally connects macroscopic observables even though they can be very challenging to calculate independently, and is largely independent of the nature of the thermal bath. Experimentally, we leverage the controllability and isolation of an ultracold system~\cite{navon2021quantum,vale2021spectroscopic} to establish and explore this connection.

Our setup, shown in Fig.~\ref{Fig1}a, is a spin-1/2 particle embedded in a thermal bath at temperature $T=1/(\beta \kB)$~\cite{leggett1987dynamics}. 
In the cartoon, the spin is depicted as a blue arrow and the bath is in red; an external near-resonant oscillating field couples the two spin states (green). The full Hamiltonian is $\hat{H} = \hat{H}^\text{s} + \hat{H}^\text{B} + \hat{H}^\text{int}$, where the spin part reads $\hat{H}^\text{s} = -\frac{1}{2}\hbar\Delta\hat{\sigma}_z + \frac{1}{2}\hbar\Omega_0\hat{\sigma}_x$ in the rotating frame of the drive; $\Delta$ is the detuning, $\Omega_0$ is the Rabi frequency, and $\hat{\sigma}_{x,z}$ are the Pauli operators ($\hbar$ is the reduced Planck constant). We make no assumptions on $\hat{H}^\mathrm{B}$ and a minimal one for $\hat{H}^\mathrm{int}$: the spin-bath interaction does not induce spin flips.

The spin is initialized in state $\ket{\downarrow}$ (see Fig.~\ref{Fig1}a) and the coupling field is switched on at $t=0$. The dynamics of the spin is monitored with the magnetization $\mathcal{M}\equiv \braket{\hat{\sigma}_z}$, as shown in Fig. \ref{Fig1}b. At long times and for sufficiently small $\Omega_0$, the spin will thermalize, with the asymptotic magnetization $\mathcal{M}_\infty\equiv\mathcal{M}(t\rightarrow\infty)$ adopting the universal form (see \emph{Methods})
\begin{equation}
\mathcal{M}_\infty(\Delta) = \tanh\left(\frac{\beta\hbar(\Delta - \Delta_0)}{2} \right), \label{Minfinite}
\end{equation}
which is characterized by a single parameter $\Delta_0$ that depends on the spin-bath interactions~\cite{vivanco2025strongly}.

At short times and small $\Omega_0$, the dynamics is instead well captured by linear response: after a brief transient, the (normalized) transition rate from $\ket{\downarrow}$ to $\ket{\uparrow}$ approaches a constant value given by the FGR: $R_\downarrow= 1/(\pi\Omega_0^2)\diff\mathcal{M}/\diff t$~\cite{chen2025emergence}.

The quantities $\Delta_0$ and $R_\downarrow$ characterize many-body physics in seemingly disjoint regimes. The zero crossing $\Delta_0$ is the effective resonance frequency of the \emph{thermalized} driven spin. 
By contrast, $R_\downarrow(\Delta)$ is the linear-response FGR spectrum, a near-equilibrium quantity in the absence of driving. 
The central discovery of this work is that they are in fact exactly related:
\begin{equation}
\hbar\Delta_0 = -\frac{1}{\beta}\ln\!\left[\int_{-\infty}^{+\infty}\! \mathrm{d}\Delta\, 
R_\downarrow(\Delta)\,e^{-\beta\hbar\Delta}\right].
\label{correspondence}
\end{equation}
We call Eq.~\eqref{correspondence} the \emph{Kubo-Thermalization correspondence}: it provides a bridge between short-time linear response and the long-time thermal properties (Fig.~\ref{Fig1}b). 
The derivation only assumes that the system relaxes to a thermal steady state under weak drive (see \emph{Methods}). 
Remarkably, Eq.~\eqref{correspondence} is independent of the microscopic form of the bath Hamiltonian $\hat{H}^\mathrm{B}$ and of the detailed $\uparrow$--bath and $\downarrow$--bath couplings, and it generalizes to an $N$-level system.

\begin{figure}[ht]
\includegraphics[width=1\columnwidth]{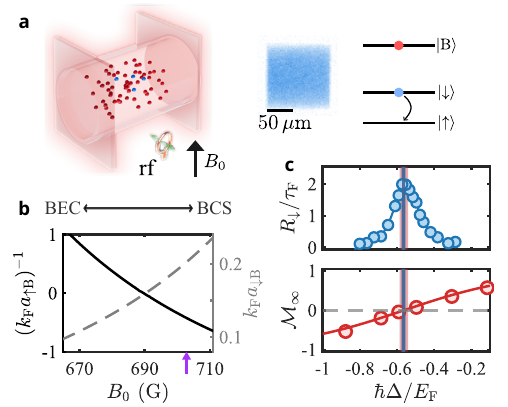}
\caption{
\textbf{Experimental platform and the Kubo-Thermalization correspondence on the BCS side.}
(\textbf{a}) (Left) Dilute spin-1/2 particles (blue spheres) consisting of $^6$Li atoms interact with a bath composed of atoms prepared in a third state $\ket{\text{B}}$ (red spheres). The system is confined in a cylindrical optical box trap, in the presence of both a static tunable magnetic field $B_0$ and a radio-frequency (rf) field. (Middle) \emph{In situ} optical density image of the spin-1/2 particles. (Right) Schematic of the level structure, showing that the bath atoms are unaffected by the rf drive that connects $\ket{\downarrow}$ and $\ket{\uparrow}$.
(\textbf{b}) Interaction strengths between two spin states and the bath atoms ($k_\text{F}a_{\uparrow\text{B}}$ and $k_\text{F} a_{\downarrow\text{B}}$) as a function of $B_0$. The BCS and BEC regimes correspond to $a_{\uparrow\text{B}}<0$ and $a_{\uparrow\text{B}}>0$, respectively. 
(\textbf{c}) (Top) Linear-response spectrum $R_\downarrow(\Delta)$ (solid blue circles) measured at $1/(\kFa)\approx-0.5$ [indicated by the purple arrow in (b)] with $\hbar\Omega_0\approx0.017\EF$ and  $t=\unit[4]{ms}$ ($\Omega_0 t\approx2.5$). (Bottom) Steady-state magnetization $\Minf(\Delta)$ at the same magnetic field with $\hbar\Omega_0\approx 0.09\EF$; in practice, $t=\unit[20]{ms}$ so that $\Omega_0 t\approx62$. The red solid line is Eq.~\eqref{Minfinite}. The vertical red band and blue band mark the positions and uncertainties of $\Delta_0$ and $\Delta_\p$, respectively.}
\label{Fig2}
\end{figure}

\begin{figure*}[ht] 
\includegraphics[width=2\columnwidth]{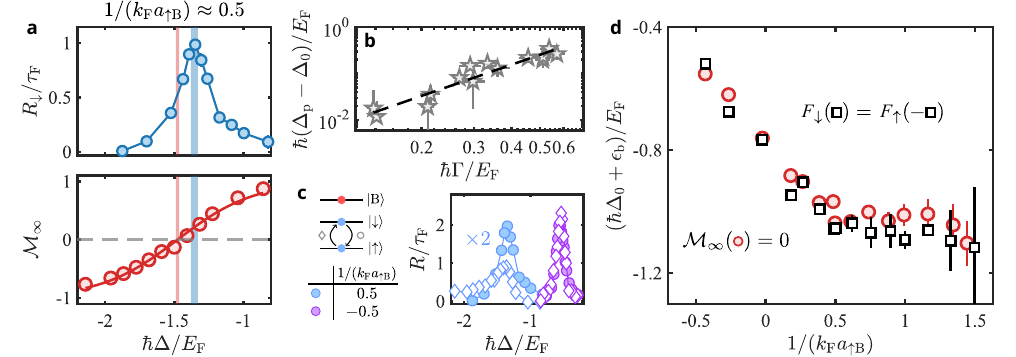}
\caption{
\textbf{Testing the Kubo-Thermalization correspondence across the BCS-BEC crossover.}
(\textbf{a}) $R_\downarrow(\Delta)$ (blue circles, top), measured with $\hbar\Omega_0\approx 0.028 \EF$ and $t=\unit[4]{ms}$ ($\Omega_0 t\approx 4$) and the steady-state magnetization $\mathcal{M}_\infty(\Delta)$ (red circles, bottom), measured with $\hbar\Omega_0\approx 0.27\EF$ and  $t=\unit[20]{ms}$ ($\Omega_0 t\approx195$). Both measurements are performed at $1/(\kFa) \approx 0.5$. The vertical red and blue bands indicate the values and uncertainties of $\Delta_0$ and $\Delta_\p$, respectively, and the red solid curve corresponds to Eq.~\eqref{Minfinite}.
(\textbf{b}) Difference $\Delta_\p-\Delta_0$ versus the full width at half maximum $\Gamma$. The black dashed line represents a power-law fit in $\log$-$\log$ scale (see text). (\textbf{c}) Examples of linear-response spectra $R_\downarrow(\Delta)$ (filled circles) and $R_\uparrow(\Delta)$ (open diamonds) on the BCS side ($1/(\kFa)\approx -0.5$, in purple) and on the BEC side ($1/(\kFa)\approx 0.5$, in blue); note that the spectra on the BEC side are multiplied by $2$ for clarity. (\textbf{d}) $\Delta_0$ versus $1/(\kFa)$ (the binding energy $-\epsilon_\mathrm{b}$ is removed on the BEC side, see text). Red circles represent experimentally determined $\Delta_0$ directly from $\mathcal{M}_\infty$, while black squares show the predicted $\Delta_0$ computed from the symmetrized Kubo-Thermalization relation (see text) using the experimentally measured $R_\uparrow(\Delta)$ and $R_\downarrow(\Delta)$ as inputs.}
\label{Fig3}
\end{figure*}

We explore the correspondence experimentally in an ultracold-atom platform: a spatially uniform gas of $^6$Li atoms confined in an optical box trap~\cite{mukherjee2017homogeneous,navon2021quantum,vivanco2025strongly} (Fig.~\ref{Fig2}a). 
The spin is encoded in two internal states of $^6$Li, denoted $\ket{\uparrow}$ and $\ket{\downarrow}$, while the bath consists of atoms in a third state $\ket{\mathrm{B}}$. 
To realize the individual-spin scenario, we prepare highly imbalanced mixtures with spin fraction $x\equiv n_{\downarrow}^{(0)}/n_{\mathrm{B}}\ll1$, where $n_{\downarrow}^{(0)}$ and $n_{\mathrm{B}}$ are the initial densities of spins and bath atoms. 
In practice, $x\lesssim0.15$, making spin--spin interactions negligible and the back-action of the spins on the bath weak. 
The bath Fermi energy is $E_\mathrm{F}\sim 2\pi\hbar\times\unit[6]{kHz}$, corresponding to a Fermi time $\tau_\mathrm{F}\equiv \hbar/E_\mathrm{F}\sim\unit[25]{\mu s}$; the temperature of the bath is $T=0.25(2)\,T_\mathrm{F}$ with $T_\mathrm{F}=E_\mathrm{F}/\kB$, and $\kB$ is Boltzmann's constant (see \emph{Methods}).

At $t=0$ we apply a radio-frequency (rf) field with Rabi frequency $\Omega_0$ and detuning $\Delta$, defined relative to the bare $\ket{\downarrow}$--$\ket{\uparrow}$ transition in the absence of the bath. 
Spin--bath interactions are set by the s-wave scattering lengths $a_{j\mathrm{B}}$ ($j=\downarrow,\uparrow$) and are tuned using a magnetic Feshbach resonance~\cite{chin2010feshbach} (Fig.~\ref{Fig2}b). 
We choose a range of bias fields $B_0$ for which the two spin states couple very differently to the bath: typically $\ket{\uparrow}$ is strongly interacting ($|\kF a_{\uparrow\mathrm{B}}|\gtrsim1$) while $\ket{\downarrow}$ is weakly interacting ($|\kF a_{\downarrow\mathrm{B}}|\ll1$), as shown in Fig.~\ref{Fig2}b. 
This configuration provides a near-ideal setting to test the correspondence.

We begin in the most tractable regime, the Bardeen--Cooper--Schrieffer (BCS) side of the Feshbach resonance ($a_{\uparrow\mathrm{B}}<0$), where $\ket{\uparrow}$--$\ket{\mathrm{B}}$ interactions produce well-defined quasiparticles known as attractive Fermi polarons~\cite{chevy2010ultra,massignan2014polarons}. At short times, we measure the linear-response spectrum $R_\downarrow(\Delta)$, normalized to the bath timescale $\tau_\F$ (top panel of Fig.~\ref{Fig2}c). 
The spectrum is narrow and nearly symmetric, with a peak at $\Delta_\mathrm{p}$ (vertical blue band). 
The shift of $\Delta_\mathrm{p}$ from zero is a direct consequence of interactions.

At long times, we measure the steady-state magnetization spectrum $\mathcal{M}_\infty(\Delta)$ reached after a long (but weak) rf pulse (bottom of Fig.~\ref{Fig2}c)~\cite{Note1}. 
We find that $\mathcal{M}_\infty(\Delta)$ follows Eq.~\eqref{Minfinite} closely (red solid line), and that $\Delta_0$ (vertical red band) coincides with the spectral peak $\Delta_\mathrm{p}$. 
This agreement is the expected limit of the Kubo-Thermalization correspondence when $R_\downarrow(\Delta)= \delta(\Delta-\Delta_\mathrm{p})$, in which case Eq.~\eqref{correspondence} reduces to $\Delta_0=\Delta_\mathrm{p}$.

We next perform a more stringent test of the correspondence in a regime where $R_\downarrow(\Delta)$ departs markedly from a delta function. 
This is realized on the Bose--Einstein condensation (BEC) side of the Feshbach resonance, $a_{\uparrow\mathrm{B}}>0$, where stronger impurity--bath coupling broadens the excitation spectrum (Fig.~\ref{Fig3}a). 
Although the steady-state magnetization spectrum still follows Eq.~\eqref{Minfinite} (bottom panel of Fig.~\ref{Fig3}a), the inferred $\Delta_0$ now clearly differs from $\Delta_\mathrm{p}$. 
To quantify this deviation, we measure $\Delta_\mathrm{p}-\Delta_0$ as a function of $a_{\uparrow\mathrm{B}}$. 
We find that $\Delta_\mathrm{p}\approx\Delta_0$ persists well into the BEC regime, up to $1/(\kF a_{\uparrow\mathrm{B}})\approx 0.3$, after which the discrepancy grows monotonically (Extended Data Fig.~\ref{EFig1} in \emph{Methods}).

Equation~\eqref{correspondence} suggests a natural origin for this deviation. 
Indeed, for a spectrum with finite full width at half maximum $\Gamma$, Eq.~\eqref{correspondence} generally implies $\Delta_\mathrm{p}\neq\Delta_0$.  We verify this by plotting the deviation $\Delta_\p-\Delta_0$ versus $\Gamma$, reconstructed by varying $\kFa$ (see Fig.~\ref{Fig3}b and Extended Data Fig.~\ref{EFig1} in \emph{Methods}). Interestingly, the fitted scaling $\Delta_\p-\Delta_0\propto \Gamma^{2.2(2)}$ (dashed line) is close to the prediction $\Delta_\p-\Delta_0\propto \Gamma^{2}$ from inserting a Gaussian ansatz $R_\downarrow (\Delta)=\frac{1}{\sqrt{2\pi}\gamma} \exp\left(-\frac{(\Delta-\Delta_\p)^2}{2\gamma^2}\right)$ with $\gamma=\Gamma/(2\sqrt{2\ln2})$ into Eq.~\eqref{correspondence}.

A full test of the Kubo-Thermalization correspondence requires determining both sides of Eq.~\eqref{correspondence}. 
In practice, directly evaluating Eq.~\eqref{correspondence} is challenging because the $\Delta<0$ tail of $R_\downarrow(\Delta)$ is exponentially amplified, demanding prohibitively high signal-to-noise. 
We circumvent this by deriving a symmetrized form of the correspondence $F_\downarrow(\Delta_0)=F_\uparrow(-\Delta_0)$, where  $F_j(z)\equiv\int_{z}^\infty\diff\Delta~R_j(\pm\Delta)\Big(e^{-\beta\hbar(\Delta-z)}-1\Big)$, $R_\uparrow(\Delta)$ is the spectrum measured when the spin is initialized in $\ket{\uparrow}$, and the plus (resp. minus) sign applies to $j=\downarrow$ (resp. $\uparrow$); see \emph{Methods} for the derivation. 
This symmetrized expression avoids exponential amplification of the tails in both $R_\downarrow$ and $R_\uparrow$.

In Fig.~\ref{Fig3}c, we show representative spectra $R_\downarrow$ (filled circles) and $R_\uparrow$ (open diamonds) for two values of $\kF a_{\uparrow\mathrm{B}}$. 
As shown in Fig.~\ref{Fig3}d, $\Delta_0$ extracted from $\mathcal{M}_\infty(\Delta)$ (red circles) agrees closely with $\Delta_0$ obtained from the symmetrized correspondence (black squares) across the BCS-BEC crossover; to make the comparison more demanding, we removed the binding energy $-\epsilon_\mathrm{b}\equiv -\hbar^2/(ma_{\uparrow\text{B}}^2)$ on the BEC side. The agreement is particularly striking because, in this strongly correlated regime, calculating $\Delta_0$ and $R_{\downarrow,\uparrow}$ poses serious theoretical challenges~\cite{vivanco2025strongly,PhysRevLett.133.083403,PhysRevA.109.023302}, so that no reliable predictions exist for the data in Fig.~\ref{Fig3}d. 
As an independent consistency check of thermalization, we measure the susceptibility
$\chi\equiv (\partial \mathcal{M}_\infty/\partial (\hbar\Delta))|_{\Delta=\Delta_0}$ and find that it is constant across interactions and equal to $\beta/2$, with $\beta$ independently obtained from time-of-flight thermometry of the bath (Extended Data Fig.~\ref{EFig2} in \emph{Methods}). 
This verifies that the driven spin equilibrates to the bath temperature. 
Together, Fig.~\ref{Fig3}d and Extended Data Fig.~\ref{EFig2} provide a complete experimental confirmation of the Kubo-Thermalization correspondence.

\begin{figure}[ht]
\includegraphics[width=1\columnwidth]{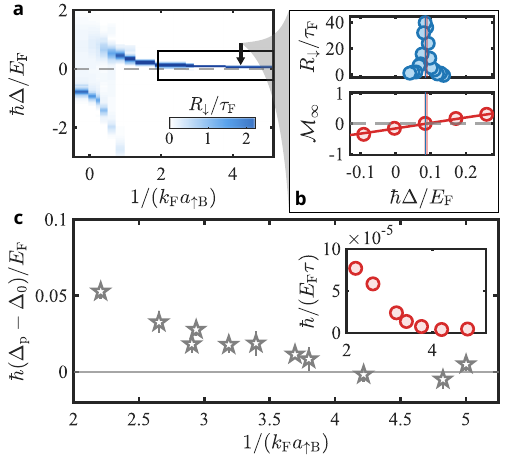}
\caption{
\textbf{The Kubo-Thermalization correspondence for the metastable (repulsive) branch.}
(\textbf{a}) Intensity map of $R_\downarrow(\Delta)$ across different interaction strengths $1/(\kF a_{\uparrow\text{B}})$. (\textbf{b}) Top panel: Linear-response spectrum $R_\downarrow(\Delta)$ (blue circles), measured with $\hbar\Omega_0\approx0.003E_\text{F}$ and $t=\unit[12]{ms}$ ($\Omega_0 t\approx 1.4$). Lower panel: steady-state magnetization $\mathcal{M}_\infty(\Delta)$ (red circles) measured with $\hbar\Omega_0\approx0.27\EF$ and $t=\unit[200]{ms}$ ($\Omega_0 t\approx 1950$). Both measurements are performed at $1/(\kF a_{\uparrow\text{B}})\approx 4.2$, shown for detunings near the repulsive polaron energy. Vertical red and blue bands mark the values and uncertainties of $\Delta_0$ and $\Delta_\p$, respectively.  
(\textbf{c}) Deviation $\Delta_\p-\Delta_0$ for the repulsive branch in the regime $1/(k_\text{F}a_{\uparrow\text{B}})\gg1$. The gray horizontal line indicates $0$. Inset: Lifetime $\tau$ of the repulsive branch.}
\label{Fig4}
\end{figure}

Finally, we show that the Kubo-Thermalization correspondence remains valid on a metastable branch. 
To this end, we focus on the regime $1/(\kF a_{\uparrow\mathrm{B}})\gg 1$, where a weakly repulsive Fermi polaron can be long-lived despite lying above the Feshbach-molecule ground state~\cite{cui2010stability,scazza2022repulsive}. 
In this limit the spectrum exhibits two branches (Fig.~\ref{Fig4}a): a metastable repulsive-polaron feature at $\Delta>0$ (highlighted by the rectangle) and an attractive molecular feature at $\Delta<0$.
Restricted to the repulsive sector, $R_\downarrow$ is sharply peaked at the repulsive-polaron energy so that we expect $\Delta_0\approx\Delta_\p$.
In Fig.~\ref{Fig4}b, we compare $R_\downarrow(\Delta)$ (blue circles) and $\mathcal{M}_\infty(\Delta)$ (red circles) near the repulsive-polaron energy for $1/(\kF a_{\uparrow\mathrm{B}})\approx 4.2$. 
The extracted $\Delta_0$ coincides with $\Delta_\mathrm{p}$, consistent with the correspondence.

This measurement is delicate because it must satisfy competing constraints: interactions must be strong enough to enable thermalization under driving, yet weak enough to suppress coupling between the repulsive and attractive branches. 
Consistent with this picture, Fig.~\ref{Fig4}c shows that the correspondence holds only within an intermediate interaction window. 
For smaller $\kF a_{\uparrow\mathrm{B}}$, thermalization becomes too slow and is ultimately limited by experimental timescales (\emph{e.g.} the vacuum-limited lifetime). 
For larger $\kF a_{\uparrow\mathrm{B}}$, the repulsive branch rapidly decays into the lower branch (which is related to losses in the three-component mixture~\cite{schumacher2026observation}). 
We confirm this interpretation by measuring the impurity lifetime $\tau$ versus $1/(\kF a_{\uparrow\mathrm{B}})$ (inset of Fig.~\ref{Fig4}c): the agreement $\Delta_0\approx\Delta_\mathrm{p}$ is obtained when thermalization occurs faster than $\tau$. 
These results show that the Kubo-Thermalization correspondence can apply even when equilibration is restricted to a sector of the Hilbert space on relevant timescales.

In summary, this work has uncovered a fundamental and rigorous correspondence between short- and long-time many-body quantum dynamics. We established this result experimentally using tunable ultracold fermions over a wide range of interactions. Because this correspondence is general, it could find applications in other systems such as NMR~\cite{du2024single}, trapped ions~\cite{monroe2021programmable}, and Rydberg atom arrays~\cite{browaeys2020many}, where similar quantum spin dynamics take place, thereby opening a new pathway to understanding quantum thermalization and its universal signatures in non-equilibrium dynamics.

\textbf{Acknowledgment.} We thank Franklin Vivanco for contributions in the early stages of this project. We thank Xiaoling Cui, Qi Gu, and Tiangang Zhou for helpful discussions, and Nathan Apfel for comments on the manuscript. N.N. acknowledges support from the David and Lucile Packard Foundation, and the Alfred P. Sloan Foundation. H.Z. acknowledges support from the National Key Research and Development Program of China (Grant No. 2023YFA1406702), and the National Natural Science Foundation of China (Grant Nos. 12488301 and U23A6004). P.Z. acknowledges support from the National Natural Science Foundation of China (Grant No. 12374477), and the Shanghai Rising-Star Program (Grant No. 24QA2700300).

\newpage
\clearpage
\setcounter{figure}{0} 
\setcounter{equation}{0}

\renewcommand\theequation{S\arabic{equation}} 
\renewcommand\thefigure{\arabic{figure}} 
\renewcommand{\figurename}[1]{Extended Data Fig.~}

\titlespacing\subsection{0pt}{11pt}{11pt}

\section{Methods}

\subsection{Preparation of highly imbalanced uniform Fermi gases}
We prepare an incoherent mixture of the first and third lowest Zeeman sublevels of $^{6}$Li, denoted $\ket{\uparrow}$ and $\ket{\B}$, in a red-detuned optical dipole trap. The impurity spin-$1/2$ states are encoded in the internal states $\ket{\uparrow}\equiv\ket{F=1/2,m_F=+1/2}$ and $\ket{\downarrow}\equiv\ket{F=1/2,m_F=-1/2}$, while the bath atoms occupy the state $\ket{\B}\equiv\ket{F=3/2,m_F=-3/2}$ (the labels $\ket{F,m_F}$ refer to the  corresponding states in the low-field basis).
Following procedures similar to Ref.~\cite{vivanco2025strongly}, we prepare a highly imbalanced mixture of states $\ket{\downarrow}$ and $\ket{\B}$ in a cylindrical optical box trap, with initial impurity fraction $x=n_{\downarrow}^{(0)}/n_{\B}\lesssim0.15$, at a magnetic field of $B\approx\unit[700]{G}$. At this field, the impurity--bath interaction is weak for $\ket{\downarrow}$, with $\kF a_{\downarrow \B}\approx0.2$. The gas is held for \unit[400]{ms} to equilibrate, after which the magnetic field is ramped to its final value $B_0$ and the system is allowed to equilibrate for an additional \unit[1]{s} before applying the rf drive. 
To initialize impurities in the $\ket{\uparrow}$ state, we apply a resonant \unit[12]{$\mu$s} rf $\pi$ pulse at \unit[700]{G} (where $1/(\kFa)\approx-0.5$) to transfer all impurities from $\ket{\downarrow}$ to $\ket{\uparrow}$, and then repeat the same sequence.

\subsection{Measurement of $\Delta_\p-\Delta_0$ versus interaction strength}
\begin{figure}[ht] 
\includegraphics[width=1\columnwidth]{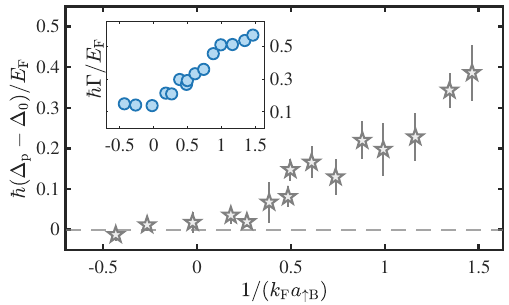}
\caption{
\textbf{Difference $\Delta_\p-\Delta_0$ as a function of $1/(\kFa)$.} Inset:
$\hbar\Gamma/E_\text{F}$ extracted from a Lorentzian fit to $R_\downarrow(\Delta)$ versus $1/(k_\text{F}a_{\uparrow\text{B}})$.}
\label{EFig1}
\end{figure}

In Extended Data Fig.~\ref{EFig1}, we show $\Delta_\p-\Delta_0$ as a function of $1/(\kFa)$. We also show the spectral width $\Gamma$ (inset) extracted by fitting a Lorentzian function to $R_\downarrow(\Delta)$; Gaussian fits give consistent results within error bars.

\subsection{Measurement of $\chi$}
\begin{figure}[ht] 
\includegraphics[width=1\columnwidth]{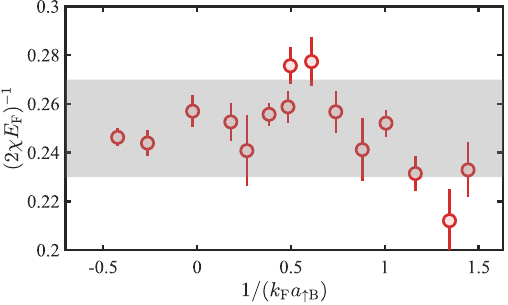}
\caption{
\textbf{Verification of thermalization.} Susceptibility $\chi\equiv (\partial \mathcal{M}_\infty/\partial (\hbar\Delta))|_{\Delta=\Delta_0}$ as a function of  $1/(k_\text{F}a_{\uparrow\text{B}})$. The gray band is determined from the bath temperature $T/T_\mathrm{F}=0.25(2)$ (which is measured by time of flight of the bath component).}
\label{EFig2}
\end{figure}

In Extended Data Fig.~\ref{EFig2}, we show the slope $\chi\equiv(\partial \mathcal{M}_\infty/\partial(\hbar\Delta))|_{\Delta=\Delta_0}$ as a function of $1/(\kFa)$. The bath temperature $T=0.25(2)T_\F$ (gray band) is extracted by time-of-flight measurement of the bath atoms.

\subsection{Derivation of the Kubo-Thermalization correspondence} We consider a spin-1/2 impurity that is driven near resonance by an external field, and coupled to a large unspecified thermal bath with temperature $T=1/\beta$ (we take $\hbar=\kB=1$ in this section). The total Hilbert space is $\mathcal H^\text{tot}=\mathcal H^\text{s}\otimes \mathcal H^\text{res}$, where $\mathcal H^\text{s}$ is the two-dimensional Hilbert space of a single spin $1/2$, and $\mathcal H^\text{res}$ is the Hilbert space of all other degrees of freedom in the problem. The total Hamiltonian is
\begin{equation}\label{gen H}
    \hat{H} = \hat{H}^\text{s} + \hat{H}^\text{B} + \hat{H}^\text{int},
\end{equation}
where the spin part (acting on $\mathcal H^\text{s}$) in the rotating frame of the drive is 
\begin{equation}
    \hat{H}^\text{s} = -\frac{1}{2}\Delta\hat{\sigma}_z + \frac{1}{2}\Omega_0\hat{\sigma}_x,
\end{equation}
where $\Delta$ and $\Omega_0$ are the detuning and Rabi frequency of the drive. Here, $\hat{\sigma}_{x,z}$ denote the Pauli operators. 
The bath Hamiltonian $\hat{H}^\text{B}$ acts on (a subset of) $\cal H^\text{res}$ and is assumed to be time independent but otherwise unspecified. The spin-bath interaction acts on $\cal H^\text{tot}$ and is assumed to take the form
\begin{equation}\label{Hint}
    \hat{H}^\text{int} = \hat{ n}_\uparrow \hat{O}_\uparrow + \hat{n}_\downarrow \hat{O}_\downarrow.
\end{equation}
The operators $\hat{O}_{\uparrow,\downarrow}$ act on $\mathcal H^\text{res}$, and $\hat{ n}_{\uparrow,\downarrow} = (1\pm\hat{\sigma}_z)/2$ are the spin projection operators for the impurity acting on $\mathcal H^\text{s}$. This form for $\hat H^\text{int}$ covers both cases of immobile and mobile impurities. Indeed, for an immobile impurity $\mathcal H^\text{res}$ coincides with $\mathcal H^\text{B}$~\cite{knap2013dissipative}, the Hilbert space of the bath. For a mobile impurity, $\mathcal H^\text{res}=\mathcal H^\text{B}\otimes\mathcal H^\text{sp}$ also includes the spatial degrees of freedom of the impurity. In that case, the energy associated with these other degrees of freedom (\emph{e.g.} the kinetic energy) can be absorbed in the definition of $\hat{O}_{\uparrow,\downarrow}$ so that the form of Eq.~\eqref{gen H} still works. We only require that the impurity-bath interactions do not flip the spin of the impurity.

We initialize the spin in either $\ket{\uparrow}$ or $\ket\downarrow$ and turn on the external drive at $t=0$. Our primary observable here is the magnetization $\mathcal{M}\equiv\braket{\hat\sigma_z}$. The key assumption of the following derivation is that the weakly driven spin thermalizes in the long-time limit with the bath, at temperature $T$. For sufficiently small $\Omega_0$, the asymptotic magnetization $\Minf\equiv\mathcal{M}(t\rightarrow\infty)$ takes the form: 
\begin{align}
   & \Minf= \frac{\text{tr}(e^{-\beta\hat{H}}\hat{\sigma}_z)}{\text{tr}(e^{-\beta\hat{H}})} \nonumber \\
    &= \frac{e^{\beta\frac{\Delta}{2}}\,\text{tr}_\text{res}(e^{-\beta(\hat{H}^\text{B}+\hat{O}_\uparrow)}) - e^{-\beta \frac{\Delta}{2}}\,\text{tr}_\text{res}(e^{-\beta(\hat{H}^\text{B}+\hat{O}_\downarrow)})}{e^{\beta\frac{\Delta}{2}}\,\text{tr}_\text{res}(e^{-\beta(\hat{H}^\text{B}+\hat{O}_\uparrow)}) + e^{-\beta\frac{\Delta}{2}}\,\text{tr}_\text{res}(e^{-\beta(\hat{H}^\text{B}+\hat{O}_\downarrow)})}, \label{thermalizedM}
\end{align}
where $\text{tr}$ and $\text{tr}_\text{res}$ respectively denote the traces over $\mathcal H^\text{tot}$ and $\mathcal H^\text{res}$.

Defining the zero crossing $\Delta_0$ as
\begin{equation}
\label{eq:detuning}
    \Delta_0 \equiv \frac{1}{\beta}\ln\frac{\text{tr}_\text{res}(e^{-\beta(\hat{H}^\text{B}+\hat{O}_\downarrow)})}{\text{tr}_\text{res}(e^{-\beta(\hat{H}^\text{B}+\hat{O}_\uparrow)})},
\end{equation}
Eq.~\eqref{thermalizedM} simplifies to the universal form Eq.~\eqref{Minfinite}:
\begin{equation}
\label{free}
    \mathcal{M}_\infty(\Delta)= \tanh\left(\frac{\beta(\Delta - \Delta_0)}{2}\right).
\end{equation}

Next, we calculate the linear-response spectrum $R_\downarrow(\Delta)$ for a spin initialized in $\ket{\downarrow}$ and coupled to $\ket{\uparrow}$ using Fermi's Golden Rule:
\begin{align}
    & R_\downarrow(\Delta) = \sum_{\nu,\mu} p_\nu \left|\bra{\uparrow;\mu}\hat{\sigma}_x\ket{\downarrow;\nu}\right|^2 \delta(E_{\downarrow,\nu}-E_{\uparrow,\mu}) \nonumber\\
    &= \frac{1}{2\pi}\int_{-\infty}^\infty \mathrm{d}t\,e^{it\Delta} \frac{\text{tr}_\text{res}\left(e^{-\beta(\hat{H}^\text{B}+\hat{O}_{\downarrow})} e^{it(\hat{H}^\text{B}+\hat{O}_{\downarrow})} e^{-it(\hat{H}^\text{B}+\hat{O}_\uparrow)}\right)}{\text{tr}_\text{res}\left(e^{-\beta(\hat{H}^\text{B}+\hat{O}_{\downarrow})}\right)}\nonumber\\
    &= \frac{1}{2\pi}\int_{-\infty}^\infty \mathrm{d}t\,e^{it\Delta}\mathcal{R}_\downarrow(t) ,
    \label{eq:injrelation}
\end{align}
where $\ket{\downarrow;\nu}$ and $\ket{\uparrow;\mu}$ denote the eigenstates of $\hat H$ at $\Omega_0=0$, where the spin is respectively in state $\ket{\downarrow}$ and $\ket{\uparrow}$ (the indices $\nu$ and $\mu$ span the Hilbert space $\mathcal H^\text{res}$). The energy of those states are $E_{\downarrow,\nu}$ and $E_{\uparrow,\mu}$ and $p_\nu = e^{-\beta E_{\downarrow,\nu}} / \sum_{\nu'} e^{-\beta E_{\downarrow,\nu'}}$ is a Boltzmann weight. $\mathcal{R}_\downarrow(t)$ is the Fourier transform of $R_\downarrow(\Delta)$. Note that $R_\downarrow$ obeys the spectral sum rule
\begin{equation}\label{sum rule}
    \int_{-\infty}^\infty\diff\Delta ~R_\downarrow(\Delta)=1.
\end{equation}

It is important to note that $\mathcal{R}_\downarrow(t)$ is analytic on the strip $-\beta < \text{Im}\,t < 0$ and continuous on its boundaries $\text{Im}\,t=0$ and $-\beta$, provided that the spectrum of $\hat H$ is bounded from below. This allows the following analytic continuation:
\begin{equation}
\mathcal{R}_\downarrow(-i\beta)=\frac{\text{tr}_\text{res}\left(e^{-\beta(\hat{H}^\text{B}+\hat{O}_{\uparrow})}\right)}{\text{tr}_\text{res}\left(e^{-\beta(\hat{H}^\text{B}+\hat{O}_{\downarrow})}\right)}=e^{-\beta\Delta_0}.  \label{eq:tildeI}
\end{equation}

We can then reverse the Fourier transform in Eq.~\eqref{eq:injrelation} and perform the analytic continuation to obtain
\begin{equation}
  \mathcal{R}_\downarrow(-i\beta)= \int_{-\infty}^\infty\diff\Delta~R_\downarrow(\Delta)e^{-\beta\Delta}=e^{-\beta\Delta_0}.
\end{equation}
Along with Eqs.~\eqref{eq:detuning}-\eqref{free}, this yields the main result, Eq.~\eqref{correspondence}.

To derive the symmetrized form of Eq.~\eqref{correspondence}, we calculate $R_\uparrow(\Delta)$ for a spin initialized in $\ket{\uparrow}$ along the lines of Eq.~\eqref{eq:injrelation}. This allows us to derive an important relation connecting $R_\uparrow(\Delta)$ and $R_\downarrow(\Delta)$:
\begin{equation}
\begin{aligned}
   R_\uparrow(\Delta)&=\frac{1}{2\pi}\int_{-\infty}^\infty \mathrm{d}t\,e^{-it\Delta} e^{\beta\Delta_0} \\
   &\times\frac{\text{tr}_\text{res}\left(e^{-\beta(\hat{H}^\text{B} + \hat{O}_{\uparrow})} e^{it(\hat{H}^\text{B} + \hat{O}_{\uparrow})} e^{-it(\hat{H}^\text{B} + \hat{O}_{\downarrow})}\right)}{\text{tr}_\text{res}\left(e^{-\beta(\hat{H}^\text{B} + \hat{O}_{\downarrow})}\right)}\\
   &=\frac{1}{2\pi}e^{\beta\Delta_0}\int_{-\infty}^{\infty} \diff t'\,e^{(it'-\beta)\Delta}\mathcal{R}_\downarrow(t')\\
   &=e^{-\beta(\Delta-\Delta_0)} R_\downarrow(\Delta).\label{relation}
\end{aligned}
\end{equation}
The first equality utilizes the analyticity of $\mathcal{R}_\downarrow$ to shift the integration contour by $-i\beta$ and a change of variable $t' = -t - i\beta$. This result, as a manifestation of detailed balance, connects the forward and reverse transition rates of a single spin in thermal equilibrium with a bath~\cite{liu2020radio}. Finally, we find the symmetrized version of the correspondence, $F_\downarrow(\Delta_0)=F_\uparrow(-\Delta_0)$, by plugging both Eq.~\eqref{relation} and Eq.~\eqref{sum rule} into Eq.~\eqref{correspondence}.

\subsection{Generalization of the Kubo-Thermalization correspondence to an $N$-level system}
Consider the following Hamiltonian:
\begin{equation}
    \hat{H}^\text{s}=\sum_i^N E_i \ket{i}\bra{i}+\frac{1}{2}\sum_{i\neq j}\Omega_{ij}\ket{i}\bra{j},
\end{equation}
where $\ket{i}$ is the eigenstate of the system with energy $E_i$. 
The system-bath interaction takes the form of $\hat{H}^\text{int}=\sum_i^N\hat{n}_i\hat{O}_i$ with $\hat{n}_i\equiv\ket{i}\bra{i}$.

Initializing the system in one of these eigenstates $\ket{k}$, we define the magnetization to be $\mathcal{M}^k\equiv\braket{\sum_{i\neq k}\hat{n}_i-\hat{n}_k}$. At long time and sufficiently small $\Omega_{ij}$, we assume that the system will thermalize with the bath and thus the magnetization will take the form
\begin{equation}
    \mathcal{M}_\infty^k(E_k)=\tanh\left(\frac{\beta(E_k-\mathcal{E}_0^k)}{2}\right),
\end{equation}
where the generalized zero crossing $\mathcal{E}_0^k$ is defined as 
\begin{equation}
    \mathcal{E}_0^k\equiv \frac{1}{\beta}\ln\frac{\text{tr}_\text{res}(e^{-\beta(\hat{H}^\text{B}+\hat{O}_k)})}{\sum_{i\neq k}\text{tr}_\text{res}(e^{-\beta(\hat{H}^\text{B}+\hat{O}_i+E_i)})}.
\end{equation}
The generalization of the correspondence involves the individual linear-response spectra from state $\ket{k}$ to state $\ket{i\neq k}$: $R_{ki}(E_k)=1/(\pi\Omega_{ki}^2)\diff\mathcal{M}^k/\diff t$, where all $\Omega_{kj}= 0$ except for $j=i$. The generalized correspondence is
\begin{equation}
    \mathcal E_0^k = -\frac{1}{\beta} \ln\left[ \sum_{i\neq k}\int_{-\infty}^\infty\diff\mathcal{E}~R_{ki}(\mathcal{E})e^{-\beta \mathcal{E}} \right].
\end{equation}

Given its general nature, it is likely that the Kubo-Thermalization correspondence could be further generalized to various other types of observables and spin-bath interactions.

\end{document}